# Deciphering the "chemical" nature of the exotic isotopes of Hydrogen by the MC-QTAIM analysis: The positively charged Muon and the Muonic Helium as new members of the Periodic Table

Mohammad Goli and Shant Shahbazian[*]

*Faculty of Chemistry, Shahid Beheshti University, G. C. , Evin, Tehran, Iran, 19839, P.O. Box 19395-4716.*

Tel/Fax: 98-21-22431661

E-mail:
(Shant Shahbazian) chemist_shant@yahoo.com

[*] Corresponding author

**Abstract**

This report is a primarily survey on the *chemical* nature of some exotic species containing the positively charged muon and the muonic Helium, i.e., the negatively charged muon plus helium nucleus, as exotic isotopes of hydrogen, using the newly developed multi-component quantum theory of atoms in molecules (MC-QTAIM) analysis, employing ab initio non-Born-Oppenhiemer wavefunctions. Accordingly, the "atoms in molecules" analysis performed on various asymmetric exotic isotopomers of hydrogen molecule, recently detected experimentally [Science 331, 448 (2011)], demonstrates that both the exotic isotopes are capable of forming atoms in molecules and retaining the *identity* of hydrogen atom. Various derived properties of atomic basins containing muonic helium cast no doubt that apart from its short life time, it is a heavier isotope of hydrogen while the properties of basins containing the positively charged muon are more remote from those of the orthodox hydrogen basins, capable of appreciable donation of electrons as well as large charge polarization; however, with some tolerance, they may be categorized also as hydrogen basins though with a smaller electronegativity. All in all, present study also clearly demonstrates that the MC-QTAIM analysis is an efficient approach to decipher the *chemical* nature of species containing exotic constituents, hard to be elucidated by experimental and/or alternative theoretical schemes.

# 1 Introduction

The heavier isotopes of the usual hydrogen ($^1H$ or $H$), i.e. deuterium ($^2H$ or $D$) and tritium ($^3H$ or $T$), are probably historically the most famous and best known isotopes of the periodic table and have found vast applications beyond chemistry from geosciences to molecular biology.[1-4] However, there have been various evidence that the positively charged muon ($\mu^+$), one of heavier congeners of electron ($e^-$) in the group of leptons,[5] may also serve as a lighter isotope of hydrogen whereas the negatively charged muon ($\mu^-$), the antiparticle of $\mu^+$, may replace electron in an atom, orbiting around positively charged nucleus tightly.[6] Although both types of muons as elementary particles have found vast applications beyond the traditional muon catalyzed nuclear fusion reactions,[7,8] deserved to be called "muon science",[9] because of their short life time, ~$2.2\times10^{-6}$s, less was known about "muonic chemistry" and most of knowledge was limited to Muonium ($Mu$) (an exotic atom composed of $\mu^+$ and $e^-$).[6] This is not an exceptional situation since tau/position, as other members of the lepton family,[5] are also unstable prone to decaying to/annihilation with electrons, leaving "leptonic chemistry" as a curiosity with less direct relevance with real life chemistry,[10,11] but a sub-branch of studying "exotic" species.[12-15] Nowadays, the situation is being changed markedly since both high level ab initio calculations as well as accessible intense muon sources with high quality muonic flows and sensitive detectors all pave the way for a better appreciation of muonic chemistry.[16-23] Particularly, recent joint experimental and theoretical studies on the isotope effect of the reactions of muonic helium (helium nucleus/alpha particle, $He^{++}$, plus $\mu^-$) as well as $Mu$ with hydrogen molecule by Fleming, Truhlar and coworkers are promising since disclose a "direct" approach to probe muon related chemical kinetics.[24-27] Based on this and similar studies, it was proposed unequivocally that $\mu^+$, denoted as $^{1.1}H$ because of its mass,[24] and muonic Helium, denoted as $^{4.1}H$ also because of its mass,[24] are lighter and heavier isotopes of $^1H$, respectively. Accordingly, Reyes and coworkers recently demonstrated computationally that adding a $\mu^-$ to an atom, with atomic number $Z$, screens completely one unit of its nucleus charge, effectively yielding an atom with atomic number $Z-1$ while adding more $\mu^-$ triggers further "transmutation" of atom yielding lower atomic numbers.[28,29] Based on this reasoning $^{4.1}H$ is effectively a hydrogen nucleus since compared to the electrons of atoms, $\mu^-$ orbits much closer to the nucleus. Based on this background, it is tempting to ask *chemical* questions such as: How much these proposed exotic isotopes of $^1H$ may be treated as hydrogen atoms when combined with other atoms forming molecules? And, do they preserve the identity of a hydrogen atom when replacing $^1H$ in a molecule? This communication is a primary attempt to respond these and similar questions using the newly developed multi-component quantum theory of atoms in molecules (MC-QTAIM).

Recently, the orthodox quantum theory of atoms in molecules (QTAIM),[30-32] as the theory of "chemical atoms", has been extended into a new framework first termed two-component QTAIM (TC-QTAIM) and then the MC-QTAIM.[33-41] In contrast to the orthodox QTAIM that only deals with electronic wavefunctions, in which the electrons act as the sole quantum particles of a molecular system and nuclei are clamped, the MC-QTAIM deals with "multi-type" wavefunctions unfolding the AIM structure of systems with various types of quantum particle. Accordingly, this novel paradigm set the stage for the AIM analysis of the multi-component systems including non-Born-Oppenhiemer description of molecules, treating nuclei as quantum waves from the outset bypassing the clamped nucleus model as well as considering exotic molecular systems containing elementary particles such as positrons and muons. Recently, the atoms in molecules (AIM) structures of some positronic diatomic species were considered in a series of reports and the *regional positron affinity* as a novel concept was emerged from this analysis.[39-41] On the other hand, atomic basins containing isotopes are distinguishable within the context of the MC-QTAIM and in the case of $LiX$ ( $X = {}^1H, {}^2H, {}^3H$ ), it was demonstrated the predicted electronegativity trend of the hydrogen isotopes is in line with previously reported high level ab initio computations.[33,34,38]

Taking this background into account, it seems reasonable to expect that the MC-QTAIM analysis of the recently experimentally considered species (*vide infra*),[24] containing $^{1.1}H$ and $^{4.1}H$, may shed some light on above posed chemical questions. Particularly, the final intend of this report is demonstrating that atomic basins containing $^{1.1}H$ and $^{4.1}H$ are viable making them new members of the periodic table probably to be placed in the "Hydrogen box".

## 2 Computational Details

In this report, inspired by the recent aforementioned study,[24] some three and four-component diatomic like systems namely, $^{1.1}H\,^1H$, $^{1.1}H\,^2H$, $^{1.1}H\,^3H$, $^1H\,^{4.1}H$, $^2H\,^{4.1}H$, $^3H\,^{4.1}H$ as well as $^{1.1}H\,^{4.1}H$ are considered using the fully variational multi-component Hartree-Fock method (FV-MC-HF) as an ab initio methodology assuming all participants as quantum particles.[42,43] The wavefunctions were expanded using [5s:1s] floating Gaussian functions (this notation implies ten s-type floating Gaussian functions for electrons and a single s-type floating Gaussian function for each of the remaining quantum particles), and simultaneously, all of their variables, i.e., exponents, coefficients and the location of functions, were optimized in a variational procedure using a non-linear optimization algorithm assuming a closed shell singlet state for electrons.[33,34] In the case of species containing $^{4.1}H$, instead of $He^{++}$ plus $\mu^-$, a separate set of the FV-MC-HF calculations were done assuming $^{4.1}H$ as a single quantum particle with a mass equal to sum of the masses of $He^{++}$ and $\mu^-$, treating the relevant species as "Model" three-component species (denoted by prefix $M-$, e.g., $M-^{4.1}H$ or $M-^1H^{4.1}H$). The details of the developed ab initio code and associated algorithms have been released previously and are not reiterated here.[34] The used masses throughout calculations are $\mu^{\pm} = 206.76828 m_e$, $^1H = 1836.152672 m_e$, $^2H = 3670.48296 m_e$, $^3H = 5496.92152 m_e$, $He^{++} = 7294.29954 m_e$, $He^{++} + \mu^- = 7501.06782 m_e$ ($m_e$ is the electron's mass). Since the whole ab initio calculations were performed within the non-Born-Oppenhiemer paradigm, it is important to realize that the resulting wavefunctions are from the class of WF1 mainly describing electrons' motions and nuclear vibrational dynamics; the translational-rotational invariance has been neglected assuming a molecular-fixed frame from the outset.[34] The resulting wavefunctions of the three-component systems were then used for the MC-QTAIM analysis. Since the wavefunctions were fully optimized, the virial theorem is satisfied automatically, $-\langle V \rangle / \langle T \rangle = 2 \pm 10^{-6}$ as is also evident from Table 1, and no extra *ad hoc* virial scaling is required for computation of atomic/basin energies.[40] Previously, the details of the topological analysis and the numerical basin integrations as well as the developed computational procedures have been disclosed in detail and are not reiterated here.[40] The sampling of space was done with great care, to ensure the precision of numerical basin integrations; therefore, the sum of the MC-QTAIM derived basin properties for topological atoms yields the total values derived independently from the ab initio calculations. All the computed AIM properties have been introduced and discussed in detail previously.[33-41] Also, all numerical values in the text, tables and figures are in atomic units so the units of physical variables are not stressed in the rest of this paper.

## 3 Ab initio Calculations

Tables 1 and 2 compress some results of the ab initio FV-MC-HF calculations on the introduced set of muonic species as well as those of $^1H\,^2H$, $^1H\,^3H$, $^2H\,^3H$, which were considered in a previous study,[36] for comparison. First, some trends in the three-component systems are discussed and then a comparison is made with the four-component systems.

The variationally optimized exponents of the s-type Gaussians describing the positively charged particles (PCPs) in the three-component systems clearly reveal the *identity* of the involved particles regardless to their *chemical environment*, ~6.0 for $^{0.11}H$, ~22 for $^1H$, ~33 for $^2H$, ~42 for $^3H$ and ~50 for $M-^{4.1}H$; it is evident from these numbers that $^{0.11}H$ has the largest and $M-^{4.1}H$ the smallest vibrational amplitudes.[35] Also, regardless of the system studied, the computed kinetic energies of $^{0.11}H$ is always the largest whereas those of $M-^{4.1}H$ is always the smallest among the considered isotopes; since, as demonstrated previously,[35] the "physics" behind using the s-type Gaussian functions for the PCPs is the 3D isotropic harmonic oscillator, these kinetic energies are somehow a measure of vibrational zero point energies. The total energies, and the mean inter-particle distances of the PCPs, in $^{0.11}HX$ ($X=^1H,^2H,^3H,M-^{4.1}H$) and $M-^{4.1}HY$ ($Y=^{1.1}H,^1H,^2H,^3H$) series are inversely proportional to the masses of $X$ and $Y$ that need further elucidation; evidently, the mass of PCPs is quite determinative and its fingerprint is seen in all the computed properties. This mass dependence is comprehensible if one notes that a larger mass implicates a more *compactly distributed/localized* quantum distribution yielding concomitantly a more condensed electronic distribution around the PCP which elucidates qualitatively why the electronic kinetic energy elevates in the aforementioned series. On the other hand, this *cooperative contraction* of PCPs' and electrons' distributions, induced by the used self-consistent field procedure of the FV-MC-HF methodology,[42,43] also makes the origin of the rise of the electron-electron ($e-e$ entry in Table 2) and the PCP-PCP ($A-B$ entry in Table 2) destabilizing interactions as well as the associated increase of the absolute magnitude of the stabilizing electron-PCP interactions ($e-A$ and $e-B$ entries in Table 2) clear. The contraction also manifests itself in the shrinkage of the mean inter-particle distances of the PCPs in the both series upon increasing the masses of $X$ and $Y$. Roughly speaking, a more localized/massive PCP repels another PCP more strongly/effectively while attracting electrons more strongly/effectively than a less localized/lighter PCP; thus, electrons are forced to encircle massive PCP in tighter orbits acquiring large kinetic energies. Accordingly, since the virial theorem has been satisfied for the considered systems, $E_{total} = -\langle T_e \rangle - \langle T_+ \rangle$, the increase in electronic kinetic energy, $\langle T_e \rangle$, of species containing heavier particles compensates the decrease in the sum of PCPs' kinetic energies, $\langle T_+ \rangle$, resulting to an overall increase in the absolute magnitude of total energies in the two series upon increasing the masses of $X$ and $Y$. The magnitude of total electric dipole moments of these species is also directly correlated with the mass difference of the two PCPs in the both series indicating that the charge distribution is more asymmetric when the mass difference grows; however, a more detailed consideration is postponed to the subsequent section after introducing the relevant MC-QTAIM analysis of basin dipoles.

The $M-{}^{4.1}HY$ series is just a model for the real four-component ${}^{4.1}HY$ ($Y={}^{1.1}H, {}^{1}H, {}^{2}H, {}^{3}H$) series thus a comparative study is due for justifying the used model systems; at first glance, inspecting Tables 1 and 2, it seems that there are large differences between the computed properties as well as the details of wavefunctions of the two series. In order to unravel the origin of these differences, the muonic helium atom ($He^{++} + \mu^-$) was separately considered at the FV-MC-HF level employing two s-type Gaussian functions located at the same center of space; the variationally optimized exponents are 30338.2500 and 180194.0161 for $\mu^-$ and $He^{++}$, respectively, while the total energy is $-257.1439$. Contrasting these exponents with those derived independently from the FV-MC-HF calculations on $^{4.1}HY$ series reveals that the exponents of the Gaussians are virtually the same, located also at the same place in space, demonstrating the fact that the two Gaussian functions describing the muonic helium atom are almost totally unaffected by the surrounding environment in the considered molecules. Accordingly, the major part of the molecules' total energies in the four-component systems originates from the strongly stabilizing interaction of $He^{++}$ and $\mu^-$ as is also evident from the computed kinetic energies as well as stabilizing $He^{++}-\mu^-$ potential energy in Table 2; the sum of the two kinetic energies is 257.1439 virtually half of the absolute value of $He^{++}-\mu^-$ potential energy, $-514.2878$, demonstrating that the virial theorem, $-\langle T \rangle = 1/2 \langle V \rangle$, holds to a high degree of precision even for the "bonded" muonic helium atom as a "quantum subsystem". Taking into account the fact that the virial theorem is also satisfied for the four-component systems, $E_{total}^{4comp.} = -\langle T_e^4 \rangle - \langle T_Y^4 \rangle - \langle T_{\alpha+\mu} \rangle$, the energy differences of the members of $^{4.1}HY$ series with the muonic helium, $\Delta E_{total}^{4comp.} = E_{total}^{4comp.} - E_{\alpha+\mu} \approx -\langle T_e^4 \rangle - \langle T_Y^4 \rangle$, are $-1.0279\,(Y=^{0.11}H)$, $-1.0926\,(Y=^1H)$, $-1.1037\,(Y=^2H)$, $-1.1087\,(Y=^3H)$. These values are much nearer to the computed total energies of the three-component model systems while always being ~0.021 lower for each congener pair of real and model systems. To dig more into the origin of this almost constant difference, employing the virial theorem for both the four and three-component systems, the following quantity, $\Delta\Delta E_{total} = \Delta E_{total}^{4comp.} - E_{total}^{3comp.} \approx -\langle T_e^4 \rangle - \langle T_Y^4 \rangle + \langle T_e^3 \rangle + \langle T_Y^3 \rangle + \langle T_{MH}^3 \rangle$ is considered where $\langle T_{MH}^3 \rangle$ is the kinetic energy of $M-{}^{4.1}H$ as the "model" particle. Inspection of Table 2 reveals that $\langle T_Y^4 \rangle \approx \langle T_Y^3 \rangle$, thus the equation simplifies further to $\Delta\Delta E_{total} \approx (\langle T_e^4 \rangle - \langle T_e^3 \rangle) + \langle T_{MH}^3 \rangle$, while inspection of the same table demonstrates that for all systems $\langle T_e^4 \rangle - \langle T_e^3 \rangle \approx 0.031$ and $\langle T_{MH}^3 \rangle \approx 0.010$. Evidently, the increase in electronic kinetic energy, because of extreme localization of $He^{++} + \mu^-$ pair in contrast to the single model PCP ($M-{}^{4.1}H$), is the origin of the lower energies of the four-component systems in comparison to their three-component model congeners. It is timely to emphasize that the extreme localization of $He^{++} + \mu^-$ pair also manifests itself on larger destabilizing contributions of $e-e$ and $Y-(He^{++} + \mu^-)$ (sum of $Y-He^{++}$ and $Y-\mu^-$ potential energies) interactions as well as the shrinkage of mean distance between the pair ($He^{++} + \mu^-$) and $Y$ in comparison to the three-component congeners. Comparison of other energy components in Table 2 also reveals the fact that the absolute magnitude of stabilizing $Y-\mu^-$ interaction is exactly half of the destabilizing $Y-He^{++}$ interaction while the absolute magnitude of the stabilizing $e-He^{++}$ interaction is exactly twice larger than the destabilizing $e-\mu^-$ interaction; in all these cases by replacing a point charge with one/two unit of charge instead of $\mu^-/He^{++}$ the same interaction energies are exactly reproduced. Finally, the absolute magnitude of stabilizing $e-Y$ interaction in the four-component systems is slightly larger than in the three-component systems, another manifestation of the shrinkage of both electrons' as well as $Y$'s distributions in the four-component systems relative to those in the three-component systems.

All in all, it is safe to claim that for all practical proposes relevant to this report replacing a fictitious PCP, $Y$, instead of $He^{++} + \mu^-$ pair seems legitimate while for the former the associated s-type Gaussian function, instead of describing internal dynamics of the muonic helium, unfolds the vibrational dynamics of $Y$ in the $M-{}^{4.1}HY$ series of species. In the next section, only the three-component systems are used for a comparative MC-QTAIM analysis and thus the contributions of all particles, except electrons, in basin energies originate just from vibrational dynamics relegating the need for disentangling contributions arising from the internal and vibrational dynamics of the muonic helium.

## 4 MC-QTAIM Analysis

The MC-QTAIM analysis of the three-component systems begins with the topological analysis of the proper Gamma density, $\Gamma^{(3)}(\vec{q})$, deciphering critical points (CPs) of the gradient vector field of the Gamma, $\vec{\nabla}\Gamma^{(3)}(\vec{q})$, as well as the inter-atomic surfaces and atomic basins.[33,34] Subsequently, various atomic properties of AIM are computed by basin integration of property densities, $\tilde{M}(\Omega) = \int_{\Omega} dq\ \tilde{M}(q)$, taking into account the fact that all the three components are participating in shaping each property density, $\tilde{M}(q) = \sum_{n=1}^{3} M_n(q)$.[35,36,38]

The topological analysis was done using $\Gamma^{(3)}(\vec{q}) = \rho_{-}(\vec{q}) + (m_e/m_\mu)\rho_\mu(\vec{q}) + (m_e/m_X)\rho_X(\vec{q})$ as the basic field revealing AIM structure in ${}^{0.11}HX$ series while for $M-{}^{4.1}HY$ series the underlying "combined" density is $\Gamma^{(3)}(\vec{q}) = \rho_{-}(\vec{q}) + (m_e/m_{\alpha+\mu})\rho_M(\vec{q}) + (m_e/m_Y)\rho_Y(\vec{q})$. In these equations $\rho_n(\vec{q})$ stands for the one-densities of each type of quantum particles, $\rho_n(\vec{q}) = N_n \int d\tau'_n\ \Psi\Psi^*$, to be distinguished through its subscript while $N_n$ is the number of corresponding quantum particles and $d\tau'_n$ implies summing over spin variables of all quantum particles and integrating over spatial coordinates of all quantum particles except one arbitrary particle belonging to the *n*-th type of quantum particles; the subscripts minus and *M* are used for electrons and the model particle, respectively. The used wavefunction, $\Psi$, for constructing the one-densities was that derived for each species from the previously mentioned FV-MC-HF calculations upon full optimization of all its variables in the variational procedure. Table 3 and Figures 1-4 include the main results of the performed topological analysis, while for comparison, the previously analyzed $^1H\,^2H$, $^1H\,^3H$, $^2H\,^3H$, species are also included in Table 3.[36] It is evident from both the table and figures that *the topological graphs (usually called molecular graphs) of all species are characteristics of diatomic species*;[30-32] two (3, -3) CPs and a single (3, -1) CP in between. In both series of species, the results of the topological analysis are predictably mass-dependent; the amount of the Gamma and the absolute value of its Laplacian at all CPs increase upon increasing the mass of *X* and *Y*. Particularly, the amount of the Laplacian of the Gamma at the (3, -3) CPs, ~ -4 for $^{0.11}H$, ~ -14 for $^1H$, ~ -18 for $^2H$, ~ -21 for $^3H$ and ~ -25 for $M-^{4.1}H$, is a proper indicator, like the Gaussian exponents discussed previously, to infer the nature of the PCP shaping corresponding (3, -3) CP. The previously discussed mass-dependent contraction is also clearly observable from both the amount of the Laplacian of the Gamma at (3, -3) and (3, -1) CPs as well as the fact that the distance between the two (3, -3) CPs diminishes gradually in both series of species (from 1.404 in $^{0.11}H\,^1H$ to 1.371 in $^{4.1}H\,^3H$). A more detailed inspection of Table 3 reveals that for each considered species, (3, -3) CP associated with the lighter PCP is always closer to (3, -1) CP; in other words, the length of the "bond path",[30-32] associated with the lighter PCP, is smaller than that of the heavier PCP.

In a recent study,[37] the ratio: $0 < \Gamma^{(3)}((3,-1)CP)/\Gamma^{(3)}((3,-3)CP) < 1$ was introduced as a measure of *topological floppiness*; a large ratio denotes a more floppy whereas a small ratio points to a rigid topological structure. In $^{0.11}HX$ series, this ratio is larger in the case of $^{0.11}H$, ~ 0.87, whereas for $X$ it is much less, <0.66, while for $M-^{4.1}HY$ series it is smaller in the case of $M-^{4.1}H$, <0.70, in comparison to the concomitant PCPs, >0.71; as is expected, heavier PCPs are capable of shaping more rigid topological structures.[37] One may conclude that *though both $^{0.11}H$ and $M-^{4.1}H$ are competent to generate AIM, the latter forms more robust basins less prone to topological transitions and catastrophes upon external perturbations.*[30-32,37] It is instructive at this stage to unravel the role of each one-density in shaping $\Gamma^{(3)}(\vec{q})$; Table 3 and Figures 3 and 4 disclose some features of these one-densities. From the figures it is evident that the topological structure of the electronic one-densities is qualitatively equivalent to that of the Gamma, that is, two (3, -3) CPs and a single (3, -1) CP in between, whereas those of the PCPs' just reveal a single (3, -3) CP. Although the topological characteristics of the Gamma are mainly determined by the electronic one-density and the PCPs "direct" contributions upon their one-densities are less important, because of the self-consistent field procedure of the FV-MC-HF methodology,[42,43] the PCPs "indirect" role on shaping the electronic one-density is significant. The contribution of each one-density in $\Gamma^{(3)}(\vec{q})$ at the CPs reveals a more quantitative picture as is evident from Table 3; as is expected, $\rho_-(\vec{q})$ has always the dominant role at all CPs since, in contrast to the other one-densities, it is not mass scaled while $\rho_\mu(\vec{q})$ has the second rank of importance because of smaller mass of $\mu^+$ in regard to the other PCPs. At (3, -1) CPs of all the considered species, except those containing $^{0.11}H$, just electrons are contributing in $\Gamma^{(3)}(\vec{q})$ and for all practical proposes: $\Gamma^{(3)}(\vec{q}) = \rho_-(\vec{q})$. However, for species containing $^{0.11}H$ there is a very small contribution originating from $\rho_\mu(\vec{q})$; because of its smaller mass, $\mu^+$ one-density distribution is more diffuse than the other PCPs and leaks, albeit in very small amounts, even beyond its own basin.

Based on the topological analysis, it came out that all the considered species are composed of two atomic basins and Tables 4 and 5 contain some results of basin integrations yielding the properties of these AIM. The populations of PCPs demonstrate that each basin contains exclusively a single type of PCP and even for basins containing $\mu^+$, the previously observed small leakage into neighboring basin does not manifest itself in first two decimals of $\mu^+$ population; therefore, the *identity* of each basin is revealed by the corresponding "enclosed" PCP. In the case of electronic populations, $N_e(\Omega)$, asymmetric basin distributions are clearly observed, further corroborating the previous proposal that in the case of hydrogen isotopes, generally, *electronegativity increases upon the increase in the mass*.[33,38] Particularly, this asymmetry is the most pronounced in $^{0,11}HX$ series demonstrating that $\mu^+$ containing basins are much less capable of accumulating electrons within themselves; based on the idea of partial atomic charges,[36] the considered series are best illustrated as $^{0,11}H^{+\delta}X^{-\delta}$ and $M-^{4.1}H^{-\delta}Y^{+\delta}$. This population asymmetry also manifests itself in atomic volumes that are markedly smaller in the case of *muonic basins* whereas the counter basins in $^{0,11}HX$ series are distinctly "expanded" compared to congeners in muonless species. Furthermore, inspection of Table 5 demonstrates that the "contracted" muonic basins have larger polarization dipoles (*vide infra*) and thus more polarized than the other considered atomic basins. Basin energies also reveal interesting mass-dependent regularities; in line with the case of populations, each PCP in each of the considered species just contributes to the energy of a single basin, $E_{PCP}(\Omega)$, and these energies may also be used as "fingerprints" to reveal the identity of the PCP. The latter observation is easily rationalized since based on a previous analytical study,[35] it is straightforward to demonstrate that $E_{PCP}(\Omega) = -3\alpha/2m$, where $\alpha$ and $m$ are the exponent of the used s-type Gaussian function (see Table 1) and the mass of PCP, respectively; both of the mass and exponent are unique for each of PCPs thus yielding a predictable/unique energy contribution. As is evident from Table 4, for the heavier PCPs the mass increase dominates the equation and $E_{PCP}(\Omega)$ diminishes; this observation is also in line with the well known fact that the heavier isotopes of hydrogen have smaller vibrational zero-point energies.[1] On the other hand, the electronic contribution to basin energies, $E_-(\Omega)$, are clearly more varied and Figure 5 reveals that they are correlated with the electronic population; the *gross structure* of this figure is composed of three "clusters" of points demonstrating that basins with similar electronic populations are also contributing almost similarly to $E_-(\Omega)$ while "clusters" with larger electronic populations are energetically more stabilized. The *fine structure* of this figure is revealed by classifying the figure into five subsets, each composed of four basins encompassing the same PCP (each subset is depicted with a distinct geometrical object in Figure 5) and then seeking for a relationship between $N_e(\Omega)$ and $E_-(\Omega)$ in each subset. Interestingly, a simple linear equation, $E_-(\Omega_i) = a_m N_e(\Omega_i) + b_m$ ($i = 1-4$, where $i$ is an index to distinguish basins), was found to be relatively effective; $a_m, b_m$ being regression parameters (the symbol $m$ is used as a subscript to emphasize the mass dependence of the parameters), to be determined in regression procedure for each subset. The total basin energy is just the sum of the encompassed PCP's and electronic contributions,[35] $E_{total}(\Omega) = E_-(\Omega) + E_{PCP}(\Omega)$. In each species, the absolute amount of $E_-(\Omega)$ is larger for the basin encompassing the heavier PCP while the revere is true in the case of $E_{PCP}(\Omega)$ however, in this "competition" the electronic contribution dominates and the absolute value of $E_{total}(\Omega)$ is larger for the heavier particle's basin.

It is timely to uncover the patterns observed for the computed electric dipoles of the species as well as the basin contributions; in the both series of species, $^{0.11}HX$ and $M-^{4.1}HY$, Table 1 demonstrates that upon the increase in the "mass difference" between $X/Y$ and $^{0.11}H/M-^{4.1}H$, the total ab initio computed electric dipoles increase. In contrast to the populations and energies, the electric dipole of each species is decomposed to various charge transfer (*CT*) and polarization (*P*) (sometimes also called first moments) contributions, only the latter is composed of basin contributions.[35] The *CT* dipoles in Table 5 have been calculated for each species assuming the (3, -3) CPs of the PCPs' one-densities as the centers of the *local* coordinate systems,[35,36] each located in a separate basin (Figures 3 and 4), while the basin *P* dipoles have also been computed employing the same local coordinate systems. Using these local coordinate systems, the *P* dipoles of the PCP's distributions are null and just the electronic *P* dipoles contribute to the total dipole of each species composed of two basins denoted by *A* (encompassing the lighter PCP) and *B* (encompassing the heavier PCP).[35] Accordingly, the final equation relating total electric dipole to *CT* and *P* contributions is considerably simplified yielding: $\vec{d}_{total} = \vec{P}_-(\Omega_A) + \vec{P}_-(\Omega_B) + \vec{d}^{CT}$ where $\vec{d}^{CT} = (1 - N_-(\Omega_A))\vec{R}$ ($\vec{R}$ is the vector connecting the two (3, -3) CPs of the PCPs' one-densities) and $\vec{P}_-(\Omega_t) = -\int_{\Omega_t} d\vec{r}_t^- \ \vec{r}_t^- \rho_-(\vec{r}_t^-)$ ($t = A, B$ and $\vec{r}_t^-$ are vectors locating electrons from the center of the corresponding local coordinate systems).[35,36] It is evident from Table 1 that the total electric dipoles of the species in $^{0.11}HX$ series are considerably larger than the others while inspection of Table 5 reveals the fact that *CT* dipoles as well as the *P* dipoles of muonic basins are large in this series, though counteracting, revealing the origin of the aforementioned large total dipole moments. Seemingly, in this series the large charge transfer, demonstrated previously considering electronic populations, orients the two *P* dipoles in the same direction, counteracting *CT* dipole, whereas for the rest of species, the two *P* dipoles are counteracting themselves. All in all, although the orientation of the total dipole moments conforms to that of the *CT* dipoles (except from $M-^{4.1}H^3H$ with its very small total and *CT* dipoles), without taking the contribution of *P* dipoles into account, the magnitudes of total dipoles are not generally reproducible.

## 5 Conclusion

The MC-QTAIM analysis performed in this study casts no doubt that both muon and muonic helium are capable of forming their own atomic basins. Particularly, the formed AIM of muonic helium are very similar to the AIM formed by the orthodox isotopes of hydrogen thus neglecting its microsecond life time, it seems safe to claim that muonic helium behaves exactly in the same way as one expects from a heavy isotope of hydrogen. This is an interesting observation since other exotic heavy isotopes of hydrogen discovered recently,[44-47] including $^4H, ^5H, ^6H$ and $^7H$, have extremely short lifetimes, ~$10^{-22}$s, and probably not amenable to kinetic experiments similar to those performed recently on species containing muonic helium.[24] However, extrapolating present MC-QTAIM analysis to larger masses, it seems safe to claim that atomic basins containing such heavy particles are not much different from the AIM produced from the orthodox isotopes of hydrogen. The case of AIM containing $\mu^+$ seems more tricky since its associated AIM behave rather differently; though, even in this case one may yet claim that we are faced with a light isotope of hydrogen with a smaller electronegativity (a more radical approach will be placing $\mu^+$ in a new box in the periodic table before hydrogen though in a very recent authoritative review on the "chemistry" of $\mu^+$ it has been described as just the "second" radioisotope of hydrogen).[48]

Although nature does not afford us other leptons between the mass of proton and electron for further experimental studies on this mass region, the positively and negatively charged pions as well known mesons are other elementary particles to be used as proper candidates for future MC-QTAIM analysis (since they belong to the Hadron family,[5] they are not really elementary but composed of a quark and an anti-quark). Recent theoretical advances in the field of the Hadronic atoms and molecules,[49,50] systems generally composed of electrons, nuclei and various hadrons, yield a promising territory for future MC-QTAIM analysis on the species containing various elementary particles, those having a mass between the masses of electron and proton.

More MC-QTAIM analysis on larger molecules containing $\mu^+$ is also needed to firmly distinguish the similarities and dissimilarities of the AIM containing $\mu^+$ and the other hydrogen isotopes.

## Acknowledgments

The authors are grateful to Masumeh Gharabaghi and Shahin Sowlati for their detailed reading of a previous draft of this paper and helpful suggestions.

**Figure Legends**

**Fig. 1** The depicted relief map of the Gamma density for $^{0.11}H^{1}H$ (a), $^{0.11}H^{2}H$ (b), $^{0.11}H^{3}H$ (c) and $M-^{0.11}H^{4.1}H$ (d) species as well as its 2D superimposed counter map (The heavier PCP's distribution is always located in the negative side of the z-axis of the coordinate system whereas the lighter PCP's distribution is located in the positive side of the z-axis). The arrows in the 2D map are selected gradient paths of the gradient vector field of the Gamma density while blue and red dots are the (3, -3) and (3, -1) critical points, respectively. The white "threads" linking the critical points are bond paths.

**Fig. 2** The depicted relief map of the Gamma density for $M-^{1}H^{4.1}H$ (a), $M-^{2}H^{4.1}H$ (b) and $M-^{3}H^{4.1}H$ (c) species as well as its 2D superimposed counter map (The heavier PCP's distribution is always located in the negative side of the z-axis of the coordinate system whereas the lighter PCP's distribution is located in the positive side of the z-axis). The arrows in the 2D map are selected gradient paths of the gradient vector field of the Gamma density while blue and red dots are the (3, -3) and (3, -1) critical points, respectively. The white "threads" linking the critical points are bond paths.

**Fig. 3** 1D depiction of the Gamma density (black curve), the electron one-density (red curve) and the mass-scaled one-densities of PCPs (blue curve) for $^{0.11}H^{1}H$ (a), $^{0.11}H^{2}H$ (b), $^{0.11}H^{3}H$ (c) and $M-^{0.11}H^{4.1}H$ (d) species along the z-axis containing the center of Gaussian basis functions. The ridge in the mass-scaled one-density of PCPs, observed on the positive side of axis of the coordinate system with a higher height, belongs to the lighter particle whereas that observed in the negative side of axis of the coordinate system, with a smaller height, belongs to the heavier particle.

**Fig. 4** 1D depiction of the Gamma density (black curve), the electron one-density (red curve) and the mass-scaled one-densities of PCPs (blue curve) for M $M-^{1}H^{4.1}H$ (a), $M-^{2}H^{4.1}H$ (b) and $M-^{3}H^{4.1}H$ (c) species along the z-axis containing the center of Gaussian basis functions. The ridge in the mass-scaled one-density of PCPs, observed on the positive side of axis of the coordinate system with a higher height, belongs to the lighter particle whereas that observed in the negative side of axis of the coordinate system, with a smaller height, belongs to the heavier particle.

**Fig. 5** The graph of the electronic contribution of basin energies, $E_{-}(\Omega)$, versus the electronic population, $N_{e}(\Omega)$, for all the twenty derived atomic basins. For clarity, the basins are categorized into five subsets wherein each subset contains basins encompassing a distinct type of PCP. Each subset has been represented using a distinct geometrical object introduced in the above right corner as a small box.

Table 1- Some results of the ab initio FV-MC-HF calculations.

| Species A-B | Total energy | virial ratio | Exponents A | Exponents B | Dipole moment | Inter-particle Distance* |
|---|---|---|---|---|---|---|
| $\mu^+$-H | -0.9882 | 2.0000 | 5.9225 | 22.0495 | 0.073 | 1.556 |
| $\mu^+$-D | -0.9991 | 2.0000 | 5.9459 | 32.7162 | 0.083 | 1.543 |
| $\mu^+$-T | -1.0040 | 2.0000 | 5.9542 | 41.0015 | 0.088 | 1.537 |
| $\mu^+$-M | -1.0072 | 2.0000 | 5.9550 | 48.6956 | 0.090 | 1.534 |
| H-M | -1.0716 | 2.0000 | 22.5318 | 49.5255 | 0.018 | 1.450 |
| D-M | -1.0826 | 2.0000 | 33.4098 | 49.6608 | 0.007 | 1.438 |
| T-M | -1.0877 | 2.0000 | 41.8522 | 49.7218 | 0.003 | 1.432 |
| H-D** | -1.0633 | 2.0000 | 22.4844 | 33.2886 | 0.010 | 1.459 |
| H-T** | -1.0683 | 2.0000 | 22.5131 | 41.7111 | 0.015 | 1.453 |
| D-T** | -1.0793 | 2.0000 | 33.3827 | 41.8255 | 0.004 | 1.441 |
| C-(A-B) | | | C | A | B | | |
| $\mu^+$-($\mu^-$-$He^{++}$) | -258.1717 | 2.0000 | 5.9941 | 30338.2600 | 180194.1000 | 0.109 | 1.512 |
| H-($\mu^-$-$He^{++}$) | -258.2364 | 2.0000 | 22.6680 | 30338.2600 | 180194.1000 | 0.036 | 1.429 |
| D-($\mu^-$-$He^{++}$) | -258.2475 | 2.0000 | 33.6180 | 30338.2600 | 180194.1000 | 0.025 | 1.416 |
| T-($\mu^-$-$He^{++}$) | -258.2526 | 2.0000 | 42.1212 | 30338.2600 | 180194.1000 | 0.021 | 1.411 |

* This is the distance between the centers of the two s-type Gaussian functions describing PCPs.
** From Reference 36.

Table 2- Energy component analysis (both kinetic and potential energies) derived from the FV-MC-HF calculations.[*]

| Species A-B | Ke | $K_A$ | $K_B$ | e-e | A-B | e-A | e-B |
|---|---|---|---|---|---|---|---|
| $\mu^+$-H | 0.9273 | 0.0430 | 0.0180 | 0.6061 | 0.6428 | -1.5564 | -1.6690 |
| $\mu^+$-D | 0.9425 | 0.0431 | 0.0134 | 0.6102 | 0.6482 | -1.5623 | -1.6941 |
| $\mu^+$-T | 0.9496 | 0.0432 | 0.0112 | 0.6120 | 0.6506 | -1.5649 | -1.7056 |
| $\mu^+$-M | 0.9543 | 0.0432 | 0.0097 | 0.6132 | 0.6521 | -1.5665 | -1.7133 |
| H-M | 1.0433 | 0.0184 | 0.0099 | 0.6389 | 0.6897 | -1.7188 | -1.7530 |
| D-M | 1.0590 | 0.0137 | 0.0099 | 0.6431 | 0.6956 | -1.7446 | -1.7594 |
| T-M | 1.0663 | 0.0114 | 0.0099 | 0.6450 | 0.6983 | -1.7563 | -1.7623 |
| H-D | 1.0313 | 0.0184 | 0.0136 | 0.6358 | 0.6855 | -1.7142 | -1.7336 |
| H-T | 1.0385 | 0.0184 | 0.0114 | 0.6377 | 0.6881 | -1.7170 | -1.7454 |
| D-T | 1.0543 | 0.0136 | 0.0114 | 0.6419 | 0.6939 | -1.7427 | -1.7517 |

| C-(A-B) | Ke | $K_C$ | $K_A$ | $K_B$ | e-e | B-C | e-A |
|---|---|---|---|---|---|---|---|
| $\mu^+$-($\mu^-$-$He^{++}$) | 0.9844 | 0.0435 | 220.0888 | 37.0551 | 0.6207 | 1.3230 | 1.7606 |
| H-($\mu^-$-$He^{++}$) | 1.0741 | 0.0185 | 220.0888 | 37.0551 | 0.6464 | 1.4000 | 1.8012 |
| D-($\mu^-$-$He^{++}$) | 1.0899 | 0.0137 | 220.0888 | 37.0551 | 0.6506 | 1.4122 | 1.8077 |
| T-($\mu^-$-$He^{++}$) | 1.0973 | 0.0115 | 220.0888 | 37.0551 | 0.6525 | 1.4176 | 1.8107 |

|  | e-C | e-B | A-C | A-B |
|---|---|---|---|---|
| $\mu^+$-($\mu^-$-$He^{++}$) | -1.5773 | -3.5212 | -0.6615 | -514.2878 |
| H-($\mu^-$-$He^{++}$) | -1.7304 | -3.6024 | -0.7000 | -514.2878 |
| D-($\mu^-$-$He^{++}$) | -1.7563 | -3.6155 | -0.7061 | -514.2878 |
| T-($\mu^-$-$He^{++}$) | -1.7681 | -3.6214 | -0.7088 | -514.2878 |

[*] The columns with K headline stand for kinetic energies whereas other columns offer the interaction potential energy of the involved particles. Note that "e" stands for electrons while "A", "B" and "C" are used to distinguish PCPs.

Table 3- Some results of the topological analysis.[*]

| Species A-B | Distance from (3, -1) CP | | | $\Gamma^{(3)}(\vec{q})$ | | | $\nabla^2\Gamma^{(3)}(\vec{q})$ | | |
|---|---|---|---|---|---|---|---|---|---|
| | (3, -3)-A | (3, -1) | (3, -3)-B | (3, -3)-A | (3, -1) | (3, -3)-B | (3, -3)-A | (3, -1) | (3, -3)-B |
| $\mu^+$-H | 0.559 | 0 | 0.845 | 0.240 | 0.209 | 0.317 | -4.255 | -0.948 | -14.065 |
| $\mu^+$-D | 0.537 | 0 | 0.864 | 0.243 | 0.211 | 0.336 | -4.455 | -0.938 | -18.632 |
| $\mu^+$-T | 0.528 | 0 | 0.870 | 0.244 | 0.213 | 0.346 | -4.462 | -0.956 | -21.659 |
| $\mu^+$-M | 0.520 | 0 | 0.872 | 0.243 | 0.215 | 0.353 | -4.281 | -1.023 | -24.324 |
| H-M | 0.661 | 0 | 0.715 | 0.330 | 0.248 | 0.365 | -13.869 | -1.198 | -24.940 |
| D-M | 0.682 | 0 | 0.691 | 0.349 | 0.253 | 0.367 | -18.343 | -1.211 | -25.076 |
| T-M | 0.690 | 0 | 0.682 | 0.359 | 0.255 | 0.368 | -21.273 | -1.210 | -25.134 |
| H-D[**] | 0.675 | 0 | 0.704 | 0.328 | 0.244 | 0.348 | -13.829 | -1.165 | -19.221 |
| H-T[**] | 0.666 | 0 | 0.711 | 0.329 | 0.247 | 0.358 | -13.853 | -1.184 | -22.346 |
| D-T[**] | 0.687 | 0 | 0.687 | 0.349 | 0.251 | 0.360 | -18.322 | -1.197 | -22.460 |

| | $\rho_-(\vec{q})$ | | | $\rho_A(\vec{q})$ | | | $\rho_B(\vec{q})$ | | |
|---|---|---|---|---|---|---|---|---|---|
| | (3, -3)-A | (3, -1) | (3, -3)-B | (3, -3)-A | (3, -1) | (3, -3)-B | (3, -3)-A | (3, -1) | (3, -3)-B |
| $\mu^+$-H | 0.210 | 0.209 | 0.290 | 6.304 | **0.035** | 0.000 | 0.000 | 0.000 | 49.179 |
| $\mu^+$-D | 0.212 | 0.211 | 0.312 | 6.366 | **0.050** | 0.000 | 0.000 | 0.000 | 89.192 |
| $\mu^+$-T | 0.213 | 0.212 | 0.323 | 6.363 | **0.057** | 0.000 | 0.000 | 0.000 | 125.339 |
| $\mu^+$-M | 0.213 | 0.215 | 0.331 | 6.284 | **0.059** | 0.000 | 0.000 | 0.000 | 162.670 |
| H-M | 0.303 | 0.248 | 0.343 | 49.343 | 0.000 | 0.000 | 0.000 | 0.000 | 164.124 |
| D-M | 0.325 | 0.253 | 0.345 | 89.640 | 0.000 | 0.000 | 0.000 | 0.000 | 164.334 |
| T-M | 0.336 | 0.255 | 0.346 | 125.940 | 0.000 | 0.000 | 0.000 | 0.000 | 164.412 |
| H-D | 0.301 | 0.244 | 0.324 | 49.321 | 0.000 | 0.000 | 0.000 | 0.000 | 90.159 |
| H-T | 0.302 | 0.247 | 0.335 | 49.335 | 0.000 | 0.000 | 0.000 | 0.000 | 126.716 |
| D-T | 0.324 | 0.251 | 0.337 | 89.618 | 0.000 | 0.000 | 0.000 | 0.000 | 126.869 |

[*] The lower half of the table offers the amount of the one-densities of electrons and those of PCPs at the CPs of the Gamma density that their locations have been offered in the upper half of the table.
[**] From Reference 36.

Table 4- Some results of the computed basin properties.

| Species | Populations | | | | | | Atomic volumes | |
| | Basin A | | | Basin B | | | | |
| A-B | e | A | B | e | A | B | Basin A | Basin B |
|---|---|---|---|---|---|---|---|---|
| $\mu^+$-H | 0.84 | ~1.00 | 0.00 | 1.16 | ~0.00 | 1.00 | 59.8 | 76.3 |
| $\mu^+$-D | 0.82 | ~1.00 | 0.00 | 1.18 | ~0.00 | 1.00 | 58.0 | 76.8 |
| $\mu^+$-T | 0.81 | ~1.00 | 0.00 | 1.19 | ~0.00 | 1.00 | 57.2 | 77.0 |
| $\mu^+$-M | 0.80 | ~1.00 | 0.00 | 1.20 | ~0.00 | 1.00 | 56.8 | 77.1 |
| H-M | 0.97 | 1.00 | 0.00 | 1.03 | 0.00 | 1.00 | 61.6 | 64.2 |
| D-M | 0.99 | 1.00 | 0.00 | 1.01 | 0.00 | 1.00 | 62.1 | 62.5 |
| T-M | 1.00 | 1.00 | 0.00 | 1.00 | 0.00 | 1.00 | 62.2 | 61.8 |
| H-D* | 0.98 | 1.00 | 0.00 | 1.02 | 0.00 | 1.00 | 62.8 | 64.0 |
| H-T* | 0.98 | 1.00 | 0.00 | 1.02 | 0.00 | 1.00 | 62.1 | 64.1 |
| D-T* | 1.00 | 1.00 | 0.00 | 1.00 | 0.00 | 1.00 | 62.5 | 62.4 |

| | Atomic Energies** | | | | | | | |
| | Basin A | | | | Basin B | | | |
| | e | A | B | Total | e | A | B | Total |
|---|---|---|---|---|---|---|---|---|
| $\mu^+$-H | -0.382 | -0.043 | 0.000 | -0.425 | -0.546 | ~0.000 | -0.018 | -0.564 |
| $\mu^+$-D | -0.376 | -0.043 | 0.000 | -0.419 | -0.567 | ~0.000 | -0.013 | -0.580 |
| $\mu^+$-T | -0.373 | -0.043 | 0.000 | -0.416 | -0.576 | ~0.000 | -0.011 | -0.588 |
| $\mu^+$-M | -0.371 | -0.043 | 0.000 | -0.414 | -0.583 | ~0.000 | -0.010 | -0.593 |
| H-M | -0.501 | -0.018 | 0.000 | -0.520 | -0.542 | 0.000 | -0.010 | -0.552 |
| D-M | -0.523 | -0.014 | 0.000 | -0.537 | -0.536 | 0.000 | -0.010 | -0.546 |
| T-M | -0.532 | -0.011 | 0.000 | -0.544 | -0.534 | 0.000 | -0.010 | -0.544 |
| H-D* | -0.505 | -0.018 | 0.000 | -0.523 | -0.526 | 0.000 | -0.014 | -0.540 |
| H-T* | -0.503 | -0.018 | 0.000 | -0.521 | -0.536 | 0.000 | -0.011 | -0.547 |
| D-T* | -0.524 | -0.014 | 0.000 | -0.538 | -0.530 | 0.000 | -0.011 | -0.541 |

* From Reference 36.
** The contributions of each type of quantum particles as well as the total basin energies are given. The total atomic energy of each basin is the sum of contributions originating from electrons and the PCPs, denoted by *A* and *B*, e.g., $E_{total}(\Omega_A) = E_-(\Omega_A) + E_A(\Omega_A) + E_B(\Omega_A)$

Table 5- The computed charge transfer (*CT*) and first moments/polarization dipoles (*P*) of each basin.[*]

| Species A-B | *CT* dipoles | First electronic moments | |
|---|---|---|---|
| | | Basin A | Basin B |
| $\mu^+$-*H* | 0.249 | -0.160 | -0.016 |
| $\mu^+$-*D* | 0.283 | -0.170 | -0.030 |
| $\mu^+$-*T* | 0.297 | -0.173 | -0.036 |
| $\mu^+$-*M* | 0.307 | -0.175 | -0.041 |
| *H-M* | 0.045 | -0.105 | 0.078 |
| *D-M* | 0.010 | -0.095 | 0.091 |
| *T-M* | -0.002 | -0.091 | 0.096 |
| *H-D*[**] | 0.023 | -0.098 | 0.085 |
| *H-T*[**] | 0.036 | -0.102 | 0.081 |
| *D-T*[**] | 0.002 | -0.092 | 0.094 |

[*] Since the electric dipole vectors have non-zero contribution just on z-axis, going through the centers of Gaussian basis functions, just a single number, namely, the z-component of dipole vector is given in each entry.
[**] From Reference 36.

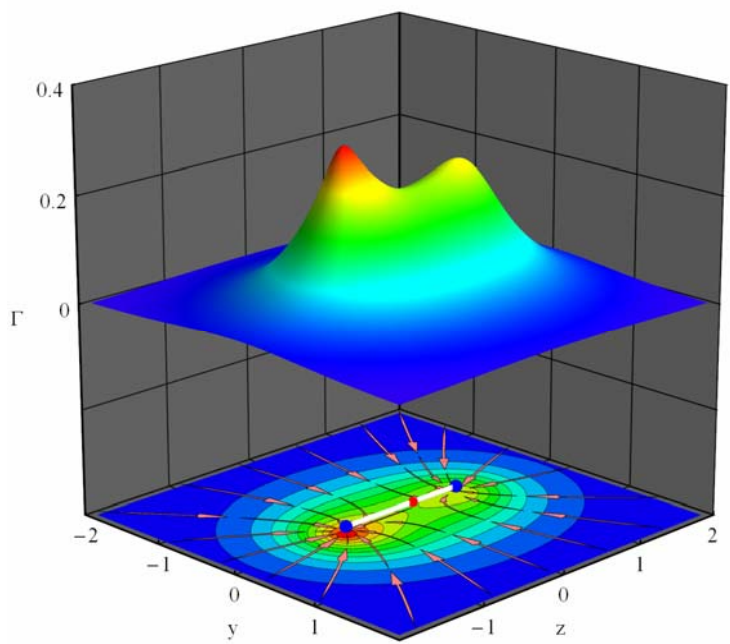 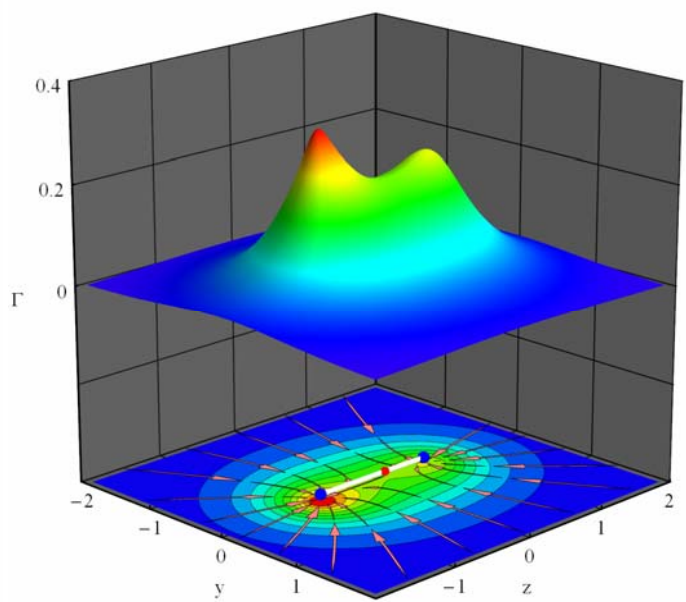

**(a)** **(b)**

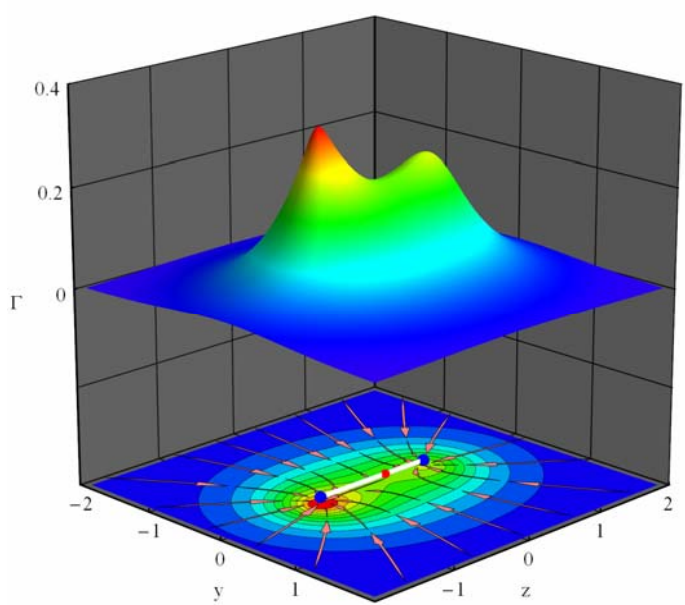 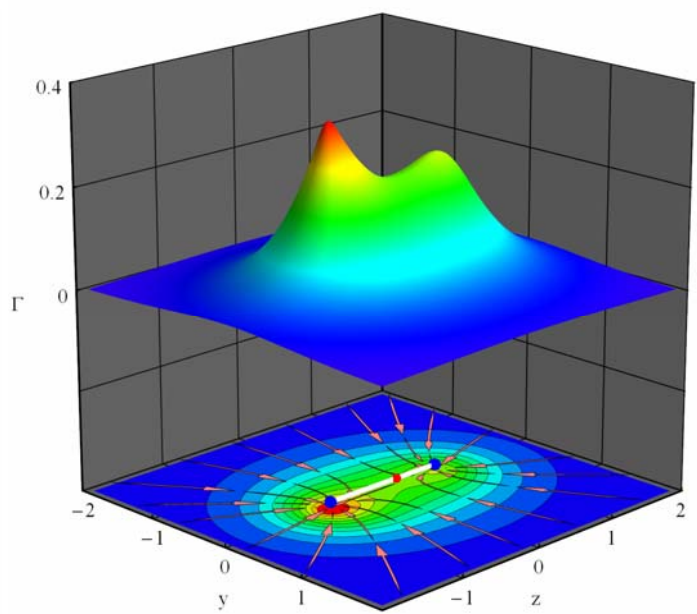

**(c)** **(d)**

**Figure-1**

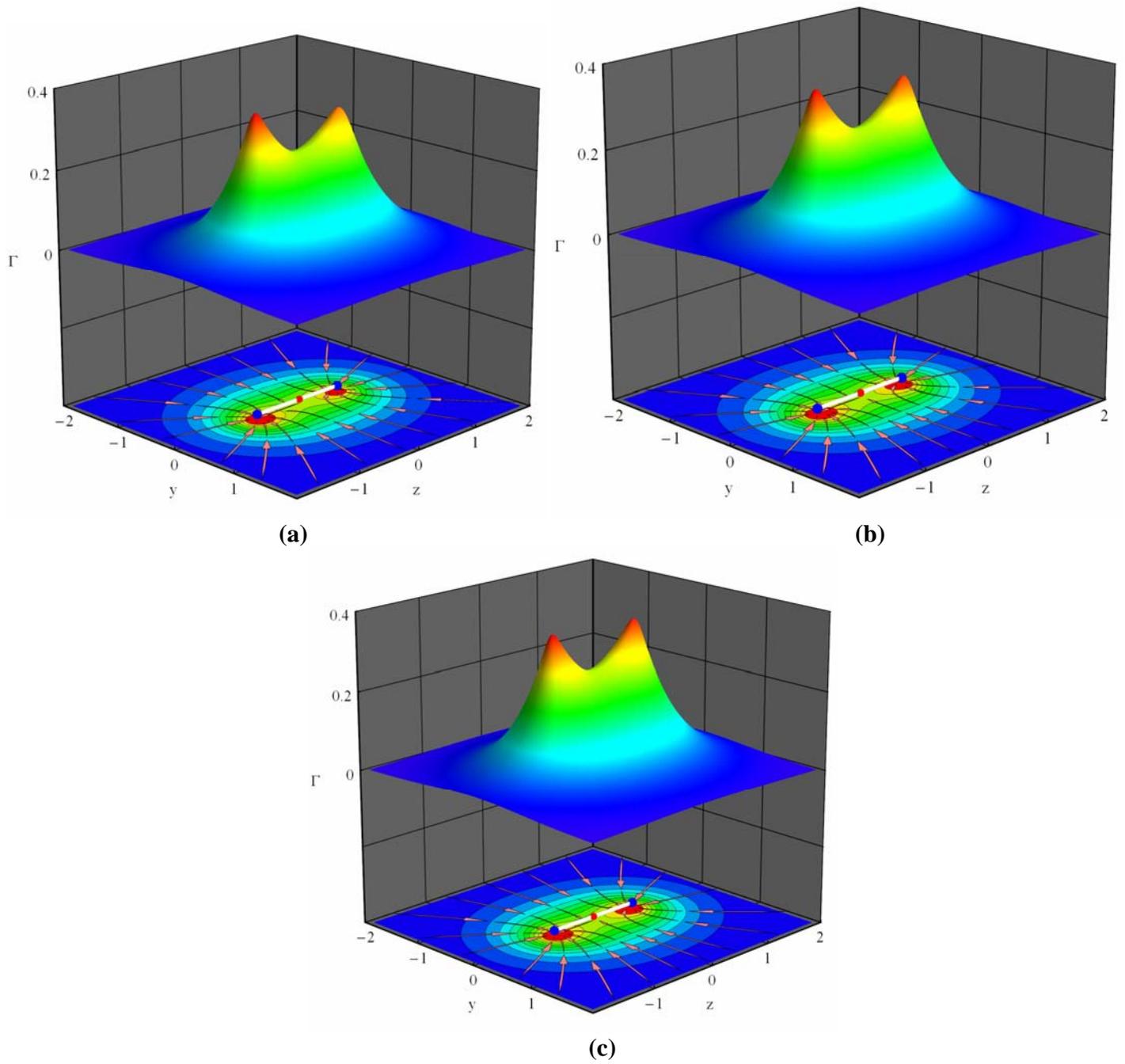

**(a)** **(b)**

**(c)**

**Figure-2**

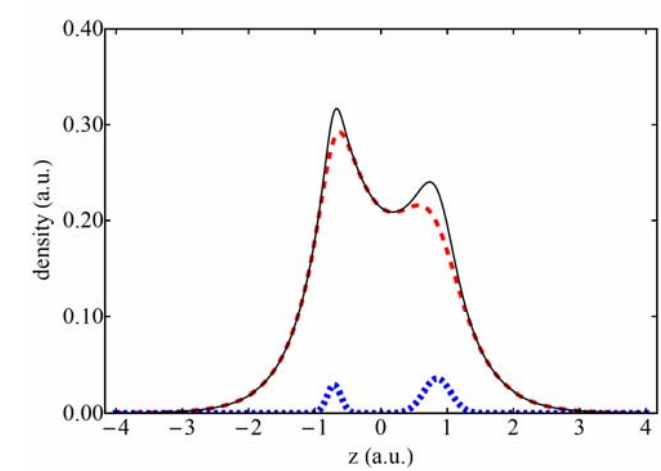 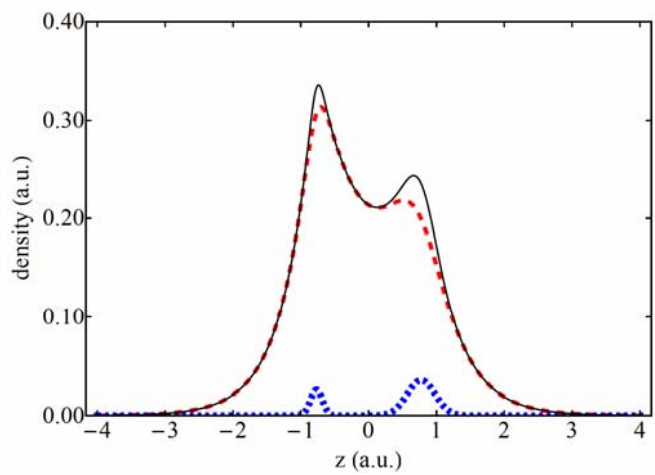

(a)　　　　　　　　　　　　　　(b)

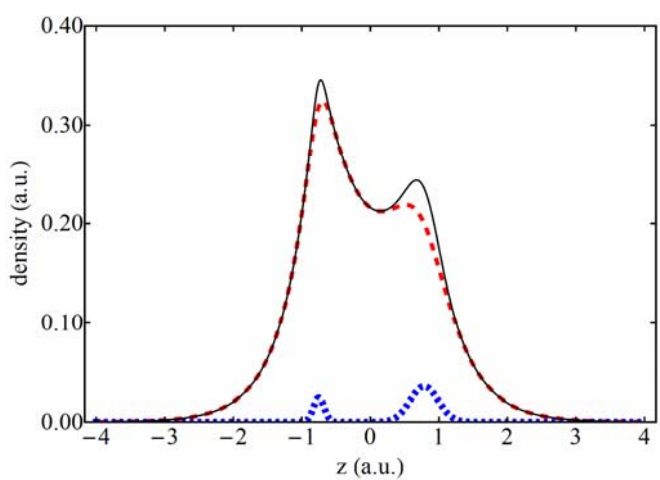 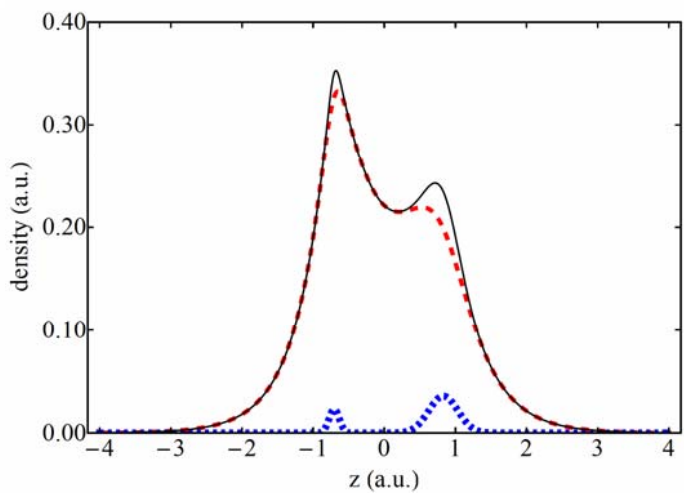

(c)　　　　　　　　　　　　　　(d)

**Figure-3**

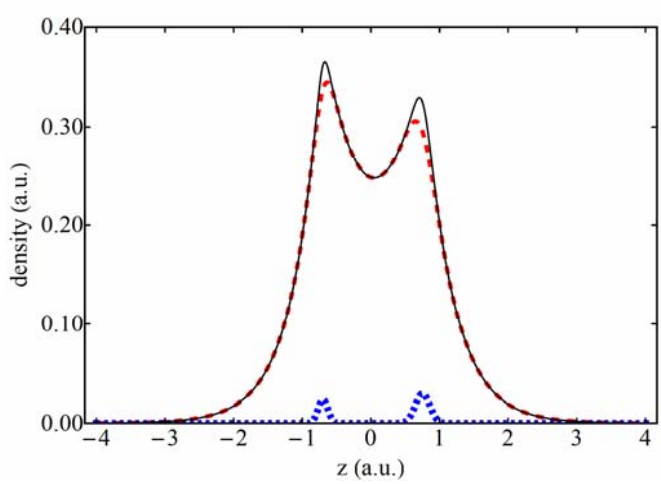

**(a)**

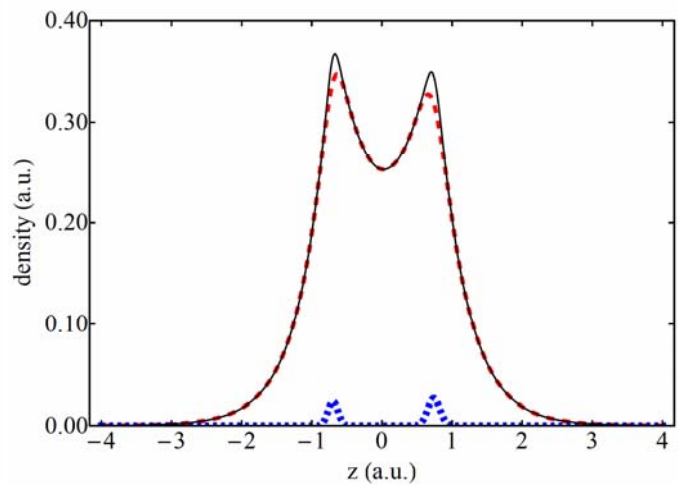

**(b)**

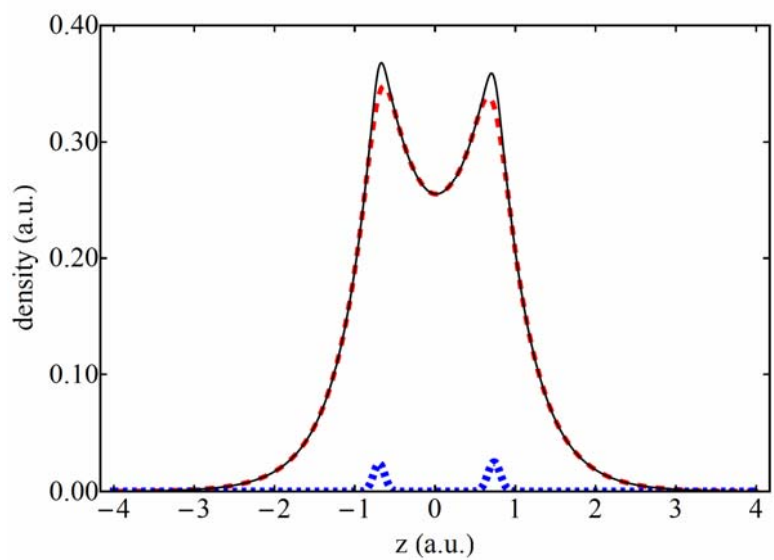

**(c)**

**Figure-4**

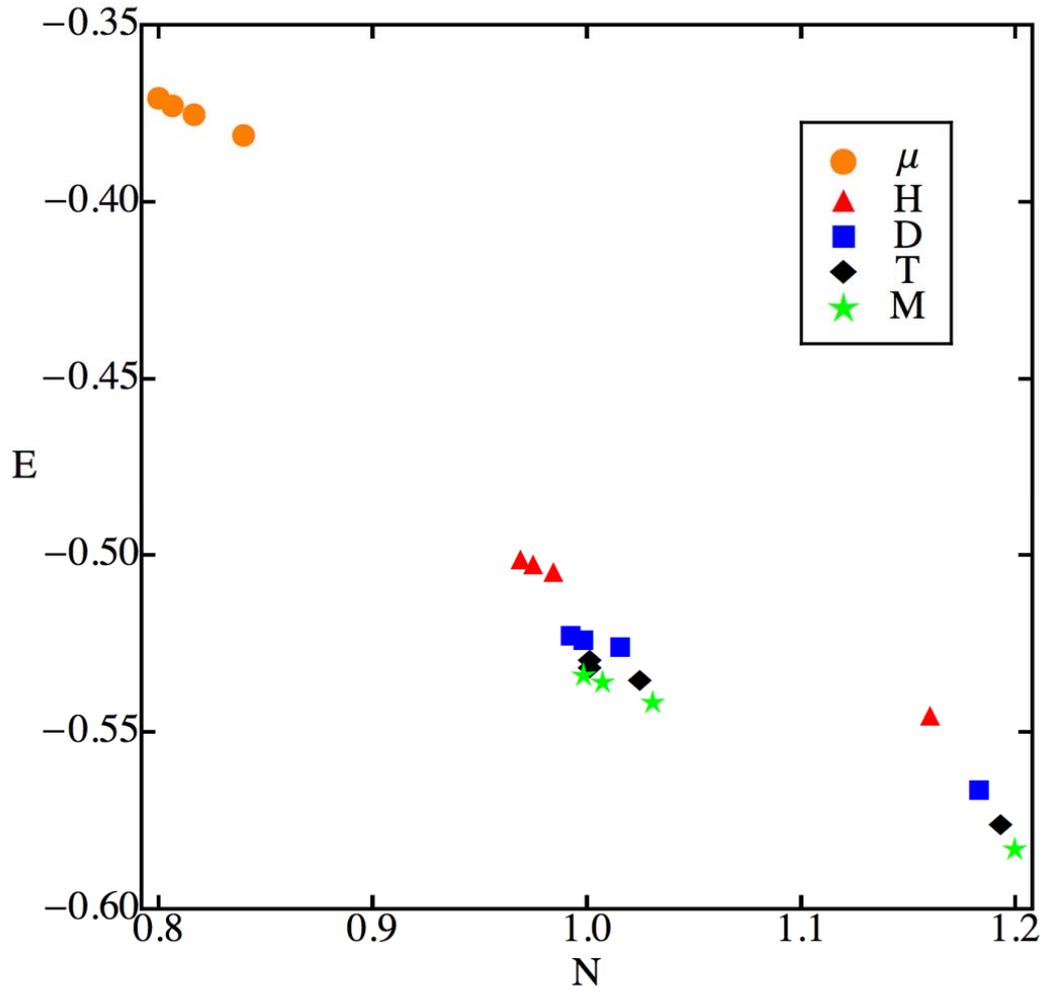

**Figure-5**